\documentclass{PoS}

\usepackage{amsmath}
\usepackage{amsfonts}
\usepackage{amssymb}
\usepackage{graphicx}
\usepackage{fontenc}
\usepackage{times}
\usepackage{mathptmx}

\newcommand{\Tr}{\mathrm{Tr}}

\newcommand{\point}{\; .}
\newcommand{\comma}{\; ,}

\newcommand{\LambdaR}{\Lambda_{\rm R}}
\newcommand{\LambdaUV}{\Lambda_{\rm UV}}
\newcommand{\SC}{\mathcal S}
\newcommand{\OC}{\mathcal O}
\newcommand{\real}{{\rm Re}\hskip 1pt}

\title{Light scalar spectrum in extra-dimensional gauge theories}

\ShortTitle{Light scalar spectrum in extra-dimensional gauge theories}

\author{\speaker{Enrico Rinaldi}\thanks{The author is supported by a
    SUPA prize studenship and a JSPS short-term fellowship.}\\
    SUPA and The Tait Institute, School of Physics and Astronomy, University of Edinburgh\\
    Edinburgh, EH9 3JZ, UK\\
    and\\
    Kobayashi-Maskawa Institute, Nagoya University,\\
    Nagoya, 464-8602, Japan\\
    E-mail: \email{e.rinaldi@sms.ed.ac.uk}}

\author{Luigi Del~Debbio\\
  SUPA and The Tait Institute, School of Physics and Astronomy, University of Edinburgh\\
  Edinburgh EH9 3JZ, UK\\
  E-mail: \email{luigi.del.debbio@ed.ac.uk}
}
\author{Alistair Hart\\
  EPCC, School of Physics and Astronomy, University of
  Edinburgh\\
  Edinburgh EH9 3JZ, UK\\
  E-mail: \email{a.hart@ed.ac.uk}
}

\abstract{The phase diagram of five-dimensional SU(2) gauge theories
  with one compactified dimension on anisotropic lattices has a rich
  structure. In this contribution we show how to control
  non-perturbatively the scale hierarchy between the cut-off and the
  compactification scale in the bare parameter space. There exists a
  set of strong bare couplings where the the five-dimensional lattice
  theory can be described by an effective four-dimensional theory with
  a scalar field in the adjoint representation.\\
  We present a detailed study of the light scalar spectrum as it
  arises from the non-perturbative dynamics of the full
  five-dimensional lattice theory. We also investigate the mixing with
  scalar glueball states in the attempt to further establish the
  extra-dimensional nature of light scalar states.\\
}

\FullConference{The 30 International Symposium on Lattice Field Theory
  - Lattice 2012,\\
  June 24-29, 2012\\
  Cairns, Australia}

\begin{document}
\section{Introduction}
\label{sect:introduction}

An elementary scalar field in any renormalisable four-dimensional
quantum field theory will get an additive mass renormalization
proportional to the second power of the ultra-violet (UV) cut-off. It
is a well known problem that is usually referred to as the \emph{naturalness
  problem} regarding the elementary scalar Higgs field in the Standard
Model.\\
Theoretical efforts have been made in order to mitigate or completely
avoid the quadratic cut-off dependence of the elementary scalar
mass. It is interesting to note that some classes of extra-dimensional
models can resolve this issue in a simple way. This is due to the
presence of an extended gauge symmetry protecting the scalar mass
from quantum correction, because in these models the scalar field is
actually one of the components of the higher-dimensional gauge
field.\\
Let us consider, for sake of definiteness, a SU($N_c$) Yang-Mills theory
in a five-dimensional Euclidean space-time defined by the following
action:
\begin{equation}
  \label{eq:continuum-action}
  \SC_E^{5d} \; = \; \int d^5x \;
  \frac{1}{2g_5^2} \: \Tr F_{MN}^2
  \comma
\end{equation}
where the field-strength tensor $F_{MN}$ is defined generalising the
four-dimensional case with $M,N=\{1,\dots,5\}$.\\
Due to its naive non-renormalizability, a UV
cut-off $\LambdaUV$ is needed and the regulated theory can be consider
as an effective physics description at energies $E \ll
\LambdaUV$ (where the details of the regularization can be neglected).
This full five-dimensional theory can be dimensionally
reduced by compactifying the extra dimension on a small circle $S^1$
of radius $R$. Below the compactification scale $\LambdaR \sim
R^{-1}$, the physics is effectively described by a four-dimensional
gauge theory with an additional scalar degree of freedom coming from
the compactified fifth component of the higher-dimensional gauge
field. This low-energy effective action can be written as
\begin{equation}
  \label{eq:4d-action}
  \SC_{\rm eff}^{4d} \; \sim \; \int d^4 x \; 
    \left[ \frac{1}{2} \Tr (F_{\mu\nu}F^{\mu\nu}) +  \Tr(D_\mu A_5)^2
    \right] \comma
\end{equation}
where $A_5$ is a scalar field in the adjoint representation of the
gauge group. The extra-dimensional origin of $A_5$ prevents its
tree-level mass to be different from zero. However, one can
calculate radiative corrections to the scalar mass
within the validity regime  $\LambdaR \ll \LambdaUV$ of
this four-dimensional (renormalisable) theory using perturbation
theory. Different approaches,
at the one and two-loop level~\cite{hosotani,cheng}, proved the
finiteness of the scalar mass and its independence of the UV
cut-off. This perturbative result is further confirmed in
Ref.~\cite{luigi} where the authors used the full
five-dimensional theory (see Eq.~\eqref{eq:continuum-action}) with two
explicit UV completions (string theory and lattice field theory) in
order to properly regulate the high-energy contributions. Although the
regulating schemes
are very different from each other, the radiative corrections to the
scalar mass of the low-energy four-dimensional theory agree with each
other when the compactification scale is well separated from the scale
where the details of the regularization appear. The renormalized
scalar mass $m_5$ is
\begin{equation}
  \label{eq:one-loop-mass}
  m_5^2 \; = \; \frac{9 g_5^2 N_c}{32 \pi^3
    R^3} \zeta (3)
  \comma
\end{equation}
where $\zeta$ is the Riemann Zeta-function. This is a very interesting
results because it seems to be universal for any UV completion of the
five-dimensional theory which respects locality and gauge-invariance.\\
It is then natural to ask what is the validity of these predictions
when the coupling constant is not small and perturbation theory is not
reliable. In particular, already in four dimensions, Yang-Mills
theories develop a mass gap non-perturbatively and confine at
low-energy, a feature that is completely missed by perturbation
theory. By studying the non-perturbative low-energy dynamics of the
theory described by Eq.~\eqref{eq:continuum-action}, we aim at
understanding the fate of Eq.~\eqref{eq:one-loop-mass}. In particular
it is important to know whether the independence of $\LambdaUV$
and the functional dependence on $\LambdaR$ are preserved in the
strong-coupling regime.\\
The results of our study are also presented in Ref.~\cite{mypaper} and
in the following we summarise and underline the main original
contributions.

\section{The lattice model}
\label{sec:lattice-setup}

Non-perturbative low-energy physics of a non-renormalizable SU($N_c$)
Yang-Mills theory can be studied using lattice Monte Carlo
simulations. The lattice acts as a gauge-invariant regulator at the UV
level and long-distance observables can be studied numerically. The
caveat is that systematic and statistical errors are introduced and
great care should be taken in order to deal with them.\\
Our lattice discretization of the action in
Eq.~\eqref{eq:continuum-action} is an anisotropic Wilson plaquette
action with two independent coupling constants:
\begin{equation}
  \label{eq:aniso-wilson}
  \SC_W \; = \; \beta_4 \sum_{x;1\leq \mu \leq
      \nu \leq 4} \left[ 1 - \frac{1}{2} \real \Tr P_{\mu\nu}(x)
    \right] +
    \beta_5 \sum_{x;1\leq \mu \leq 4}
    \left[ 1 - \frac{1}{2} \real \Tr P_{\mu 5}(x)
    \right]
    \point
\end{equation}
The four-dimensional plaquette $P_{\mu\nu}$ and the extra-dimensional
one $P_{\mu 5}$ have different couplings $\beta_4$ and $\beta_5$
allowing for two different lattice spacings to be varied
independently: $a_4$ in the four-dimensional sub-lattice and $a_5$
along the extra dimension. The physical lengths of the lattice are
determined by the following quantities: $L_4 = a_4 N_4$ and
$L_5 = 2 \pi R = a_5 N_5$; the three-dimensional volume will be $V =
(L_4)^3$, with temporal extent $L_T = L_4$ and extra-dimensional size
$L_5$.
The setup just described is well suited to study a physical system
with a compact extra dimension, because of the freedom of adjusting
both $a_5$ and $N_5$ independently of $a_4$ and $V$ in order to obtain any
possible scale separation between $\LambdaR$ and $\LambdaUV$.\\
Previous studies in the literature have explored the phase diagram of
this lattice model~\cite{ejiri,philippe,antonio} and a summary of the
most recent results can be found in Ref.~\cite{mypaper}. We focus on
the region of parameter space $(\beta_4, \beta_5, N_4, N_5)$ with
$N_5 < N_4$ and $\beta_4 < \beta_5$ where a dimensionally reduced
phase appears thanks to the compactification of the extra
dimension. In this phase, the long-distance physics of the system is
expected to be described by the four-dimensional effective theory of
Eq.~\eqref{eq:4d-action} with a light scalar particle whose mass is
perturbatively given by Eq.~\eqref{eq:one-loop-mass}. \\

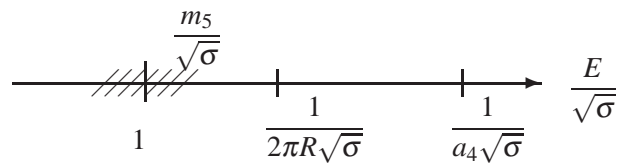
\begin{figure}[hb]
  \centering
  \begin{picture}(300,40)
    \thinlines
    \multiput(80,20)(5,0){7}{\line(1,1){10}}
    \thicklines
    \put(50,25){\vector(1,0){200}}
    \put(100,18){\line(0,1){15}}
    \put(95,0){$1$}
    \put(110,40){$\displaystyle \frac{m_5}{\sqrt{\sigma}}$}
    \put(150,20){\line(0,1){10}}
    \put(145,5){$\displaystyle \frac{1}{2\pi R \sqrt{\sigma}}$}
    \put(220,20){\line(0,1){10}}
    \put(215,5){$\displaystyle \frac{1}{a_4 \sqrt{\sigma}}$}
    \put(260,20){$\displaystyle \frac{E}{\sqrt{\sigma}}$}
  \end{picture}
  \label{fig:scales}
  \caption{The figure shows the desired separation of energy scales.
    The scales are all expressed in terms of the four-dimensional
    string tension that characterises the low-energy physics of the
    theory.}
\end{figure}
\clearpage

\section{Long-distance physics and scalar spectrum}
\label{sec:infrared-physics}

\begin{figure}[ht]
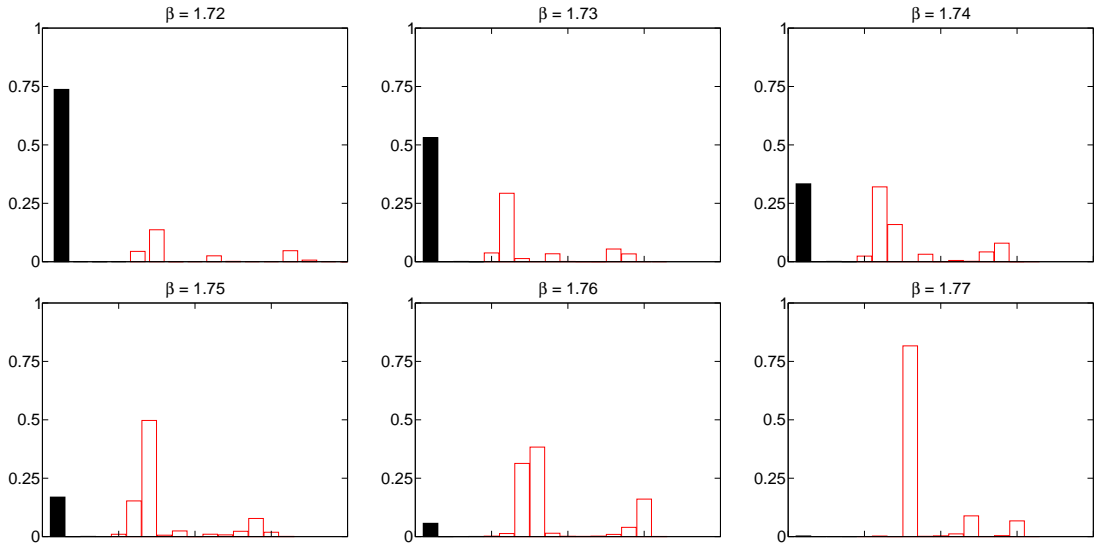

  \centering
  \begin{tabular}{ccc}
    \includegraphics[width=0.30\textwidth]{FIGS/proj_ground_1-72} &
    \includegraphics[width=0.30\textwidth]{FIGS/proj_ground_1-73} &
    \includegraphics[width=0.30\textwidth]{FIGS/proj_ground_1-74} \\
    \includegraphics[width=0.30\textwidth]{FIGS/proj_ground_1-75} &
    \includegraphics[width=0.30\textwidth]{FIGS/proj_ground_1-76} &
    \includegraphics[width=0.30\textwidth]{FIGS/proj_ground_1-77} \\
  \end{tabular}
  \label{fig:mixing-ground}
  \caption{Relative projection of the ground state onto each
    of the operators in the variational set for simulations at fixed
    $\gamma=1.5433$ and increasing $\beta$. Filled symbols
    correspond to the set of smeared scalar operators, whereas the
    open symbols refer to the smeared versions of glueball operators.}
\end{figure}

The perturbative result of Eq.~\eqref{eq:one-loop-mass} holds when
$\LambdaR \ll \LambdaUV$ and the details of the regularization can be
neglected. Also, the low-energy physics, whose energy scale is
determined by the string tension $\sqrt{\sigma}$, should be separated
from both the compactification and the cut-off scale in order for our
results to be meaningful. Therefore, the first step of our study is
to determine non-perturbatively the energy scales of the physical
system. The desired hierarchy is represented schematically in
Fig.~\ref{fig:scales}, where the string tension is used to set the
scale. The simulation's strategy is straightforward in principle:
by fixing a point in the bare lattice parameter space, we are choosing
a given separation of scales. The energy scales of the system are
determined by measuring physical observables via numerical Monte Carlo
simulations.\\
We measure $a_4\sqrt{\sigma}$ and $a_4 m_5$ by studying the large-time
exponential decay of correlators of suitable lattice operators. The
string tension is extracted from the ground state mass of correlators
of long spatial Polyakov loops. On the other hand, the scalar mass
requires us to construct operators that transform as scalars in
three dimensions and with positive parity and charge (belonging to the
$0^{++}$ irreducible representation in standard spectroscopic
notation). We distinguish between two
different kind of operators in the scalar channel: those constructed
using links $U_5(x)$ in the compact fifth direction (generically called
\emph{scalar operators} in the following) and those built using only
spatial links $U_i(x)$ (called \emph{glueball operators}).
For the first kind, we use compact
Polyakov loop operators, that is gauge-invariant combinations of
Polyakov loops winding around the extra fifth dimension.
In particular, we choose two different combinations
\begin{equation}
  \label{eq:scalar-1}
  \OC_1(t) \; = \; \sum_x \Tr[l_5(x,t)] \; ; \qquad 
  l_5(x,t)=\prod_{j=1}^{N_5} U_5(x+ja_5\hat{5},t)
  \comma
\end{equation}
and
\begin{equation}
  \label{eq:scalar-2}
  \OC_2(t) \; = \; \sum_x \Tr[\phi(x,t) \phi^\dag(x,t)] \; ; \qquad
  \phi(x,t)=\frac{l_5-l_5^\dag}{2}
  \point
\end{equation}
We average the correlators over the
extra-dimensional coordinate and correlate along the temporal coordinate.\\
For the operator in Eq.~\eqref{eq:scalar-2} we apply the smearing procedure
introduced in Ref.~\cite{irges} for a scalar Higgs field: the operator
$\phi$ is replaced by a smeared version that 
consists of a gauge-invariant combination of parallel transporters in
the three-dimensional spatial subspace. Operators $\OC_1$ and $\OC_2$
are expected to couple mainly with scalar states of extra-dimensional
nature.
\begin{figure}[hb]
  \centering
  \includegraphics[width=0.70\textwidth]{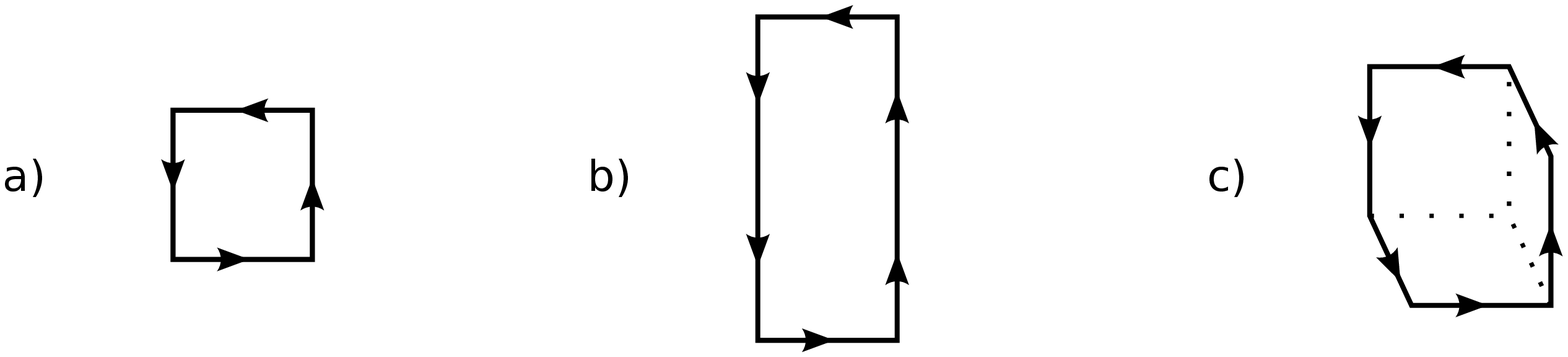}
  \label{fig:glue-path}
  \caption{Wilson loops used in the construction of glueball operators
    in the scalar channel. Each of these three operators is smeared in
    order to construct a larger variational ansatz.}
\end{figure}
To create operators coupling to scalar glueball states, instead, we
use symmetrized combinations of the spatial Wilson loops in
Fig.~\ref{fig:glue-path}. These operators
project only onto the scalar representation of the three-dimensional
cubic symmetry group are then cross-correlated together with operators
$\OC_1$ and $\OC_2$; we 
refer to these scalar glueball operators as $\OC_a$, $\OC_b$ and
$\OC_c$ built starting respectively from the path a), b) and c) in
Fig.~\ref{fig:glue-path}. The glueball operators are smeared according
to the improved blocking algorithm described in Ref.~\cite{biagio}.\\
The inclusion of glueball operators in the
study of the light scalar spectrum of this five-dimensional lattice
model is necessary in order to better
understand the fate of Eq.~\eqref{eq:one-loop-mass}. The presence of
light scalar glueballs could induce mixing with the scalar state we
expect perturbatively due to the compactification mechanism described
in Sec.~\ref{sect:introduction}. Thanks to our extended set of
operators, we can use a variational ansatz to study the scalar
spectrum and extract informations about the mixing of the ground state
with glueball operators. The variational procedure finds the linear
combination of the available operators ($\OC_1, \ldots , \OC_c$) that
best approximates the wavefunction of the lowest mass eigenstate. The
relative contribution of each operator to the mass eigenstate can be used to
infer the nature of the state: if the main contribution is due to
glueball operators, the mass extracted is likely to be associated to a
glueball state. On the other hand, a dominant contribution coming from
the scalar operators would confirm the extra-dimensional nature of the
extracted mass. This study can be performed on the lightest excitations
of the spectrum as well.\\
Due to the high computational cost of such a measurement, we employed
the full variational method on a subset of the phase diagram
explored. We choose a lattice of size $L_4=10a_4$ and $L_5=4a_5$, but
with an extended temporal direction $L_T=2L_4$ that helped us reduce
finite-size systematic effects. Then we fix a line in the parameter
space corresponding to a fixed ratio of the coupling constants
$\gamma=\sqrt{\beta_4/\beta_5} \approx 1.54$ and we simulate different
points with $\beta= \sqrt{\beta_4 \beta_5} \in \{1.72, \ldots,
1.77\}$. These points correspond to a fixed separation between the
compactification and the cut-off scale $\LambdaUV/\LambdaR \approx 2$
and increasing $\beta$ values leads to smaller extra-dimensional
radii. The perturbative expectation in this region is that the scalar
mass will increase with $\beta$.\\
\begin{figure}[hb]
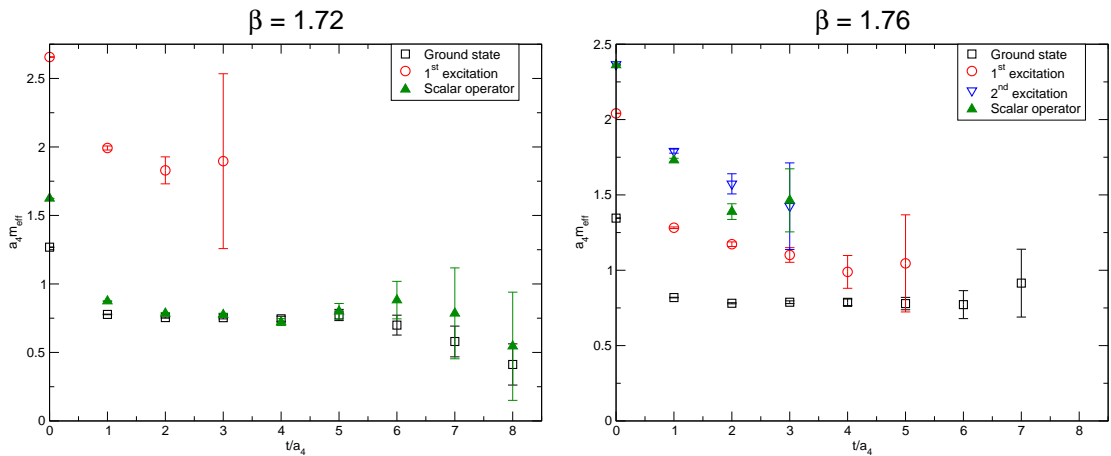

  \begin{tabular}{cc}
    \includegraphics[width=0.47\textwidth]{FIGS/eff_mass_1-72}&
    \includegraphics[width=0.47\textwidth]{FIGS/eff_mass_1-76}
  \end{tabular}
  \label{fig:plateaux-scalar}
  \caption{Example of effective mass plateaux for two different values
    of $\beta$ at fixed $\gamma=1.5433$. (left) At $\beta = 1.72$ the
    mass of the low--lying 
    scalar state obtained from the variational ansatz is compatible
    with the one we measured using only the scalar operator $\OC_2$. (right) At
    $\beta = 1.76$ the scalar operator $\OC_2$ yields a mass which is
    compatible with the second excitations of the scalar spectrum.}
\end{figure}
\clearpage
In Fig.~\ref{fig:plateaux-scalar} we show the effective masses
extracted from the variational analysis of the scalar correlators for
two different $\beta$ values. At the smallest $\beta$, the variational
ground state is the same that we obtain using scalar operators of type
$\OC_2$ and this is a first indication that the lowest propagating
state would come from the dimensional reduction of the full
five-dimensional gauge theory. On the contrary, at the largest
$\beta$, the correlators of scalar operators gives an effective mass
which is definitively heavier than the one of the variational ground
state. The scalar ground state at $\beta=1.76$ seems to be dominated
by the contributions of glueball operators. This can be verified by
looking at the relative projections of each operator onto the ground
state; the results are summarised in Fig.~\ref{fig:mixing-ground} for
all the $\beta$ values included in our study. When $\beta$ is
increased at fixed $\gamma$, the contribution to the ground state is
first dominated by scalar operators, but glueball operators start to
mix and then become dominant at large $\beta$. The emerging physical
picture is that the effective theory describing the long-distance
physics in this region of the bare coupling space, becomes more and
more four-dimensional with the lowest scalar
spectrum dominated by non-perturbative gauge excitations
(glueballs). Any sign of the extra-dimensional UV nature of the system
is lost when the excitations of the scalar field coming
from the compactification mechanism become heavy and decouple from the
low-energy dynamics. This picture is in agreement with earlier
results~\cite{philippe} that probed the four-dimensional nature of the
effective theory by carefully studying the string tension dependence
on the lattice couplings.

\section{Conclusions}
\label{sect:conclusions}

We found a region of the bare parameters of a five-dimensional lattice
model where dimensional reduction through compactification of the
extra dimension takes place. In this region
we studied the light spectrum in the scalar channel and we
determined non-perturbatively the physical energy scales $\LambdaR$
and $\LambdaUV$. When the
desired separation of scales is found (see Fig.~\ref{fig:scales}),
we expect the low-energy dynamics to be described by an effective
four-dimensional gauge theory coupled to an additional scalar
field. 
This effective description is confirmed by our measurements of the
non-perturbative spectrum. For a specific region of the parameter
space, we have shown that the lightest state
propagating in the scalar channel has a negligible mixing with
glueball states and its mass can be unambiguously identified with
$m_5$. Moreover, the functional dependence of this non-perturbative
scalar mass is found to be consistent with the perturbative prediction
of Eq.~\eqref{eq:one-loop-mass}.\\
The results of this study and of the one in Ref.~\cite{philippe} can
be considered as a good starting point for investigating the
realisation of the compactification mechanism from five to four
dimensions at the non-perturbative level.



\end{document}